\documentstyle[aps]{revtex}
\begin{document} 
\draft \title{Reflection coefficient for superresonant scattering}
\author{Soumen Basak\thanks{Electronic address: soumen@imsc.res.in}
and Parthasarathi Majumdar\thanks{On deputation at Saha Institute of
Nuclear Physics, Kolkata 700 064, India; email: 
partha@theory.saha.ernet.in}} 
\address{ The Institute Of Mathematical sciences\\ Chennai-6000113, 
INDIA} 
\maketitle
\begin{abstract}
We investigate superresonant scattering of acoustic disturbances from
a rotating acoustic black hole in the low frequency range. We derive
an expression for the reflection coefficient, exhibiting its frequency
dependence in this regime.
\end {abstract}
\bigskip 
\section{Introduction}

Progress in understanding black hole radiance, both in the
semiclassical and quantum gravitational regimes, have suffered greatly
in the past, due to the lack of experimental feedback. Black hole
radiance consists of spontaneous radiation (Hawking radiation)
\cite{haw}as well as stimulated emission (superradiance) \cite{zel},
\cite{star}, \cite{dew}, \cite{wald}. For astrophysical black holes,
the Hawking temperature is invariably smaller than the Cosmic
Microwave Background temperature. So the black hole mostly accretes
rather than radiates spontaneously. Similarly, superradiant emission
is also hard to observe from within a background of x-rays being
emitted from accretion processes. Due to these reasons physicists have
been looking for alternative physical models of gravity so that
experiments can be devised for indirect verification of phenomena that
are predicted by the semiclassical theory of gravity. These
alternative physical models are called analog models of gravity.

Among the first of such analog gravity models, Unruh's\cite{bun} was the
pioneering  idea of analog models based on fluid mechanics. The idea
of such models arose from a remarkable observation regarding sound
wave propagation in a fluid in motion. Unruh showed that if the fluid is
barotropic and inviscid, and the flow is irrotational, the equation of
motion of the  acoustic fluctuations of the velocity potential is
identical to that of a minimally coupled massless scalar field in an
effective (3+1) dimensional curved spacetime, and is given by,
\begin{equation}
\Box\Psi=\frac{1}{\sqrt{-g}}~\partial_{\mu}\left(\sqrt{-g}~
g^{\mu\nu}~\partial_{\nu}\right)\Psi=0
\label{we}
\end{equation}
where $~g_{\mu\nu}~$ is the `acoustic' metric and$~\Psi~$ is the
fluctuation of the velocity potential. The most remarkable fact about
this effective spacetime `seen' by the acoustic fluctuations is that
the acoustic metric is curved and {\it Lorentzian} in signature, even
though the fluid flow is governed by non-relativistic physics in a
flat Minkowski background spacetime. This particular feature of these
models raises the possibility of being able to mimic a gravitational
system in its kinematical sector. Since the fluid has barotropic
equation of state, the pressure does not generate vorticity. So an
initially irrotational flow will remain irrotational. Viscosity of the
fluid has been ignored in order to maintain the Lorentzian character
of the effective acoustic geometry.

In acoustic analog models of gravity, the acoustic geometry entirely
depends on the the background fluid motion. So the changes of the
acoustic geometry are also related to that of the background fluid
motion. Since the background fluid motion is governed by the Euler
equation and the continuity equation, not by Einstein's equation, it
may mimic only the kinematical aspects of gravity, and not its
dynamical ones. If the flow contains a well-defined region where it is
supersonic, that region resembles an ergoregion of a rotating black
hole, characterized by the time translation Killing vector turning
spacelike. Within this region, there may exist a surface where the
inward normal velocity of the fluid exceeds the local speed of sound
at every  point. Such a surface therefore traps {\it all} acoustic
disturbances within it, resembling an outer trapped surface, and can
therefore serve as the sonic or acoustic horizon of the analogue black
hole.

Unruh\cite{bun} has used this idea of an acoustic analogue black hole
horizon, to show that quantized acoustic perturbations (phonons) may
be emitted from it precisely like the emission of photons from a black
hole horizon via Hawking radiation. The Hawking temperature is
likewise given by the surface gravity at the sonic horizon. This is
acceptable provided we continue to ignore the backreaction of the
phonon radiation on the background metric, which is a legitimate thing
to do in a fixed-metric problem. Unruh's work has led, in the last
couple of years, to several proposals for the experimental
verification of acoustic Hawking radiation. The Hawking temperature
for typical acoustic analogues has been estimated to be in the
nano-Kelvin range; with improved experimental techniques and
equipment, this is not outside the realm of realistic possibilities in
the near future.

One slightly unsavory feature of the Unruh approach has to do with the
necessity of quantizing linearized acoustic disturbances and the
manner in which they are quantized. In a classical fluid, there is no
compelling physical reason to quantize acoustic disturbances. So
quantizing acoustic perturbations just to demonstrate Hawking
radiation from the horizon of acoustic black holes appears somewhat
artificial. In a quantum fluid like superfluid helium on the other
hand, there are phonon excitations of the fluid background itself. On
top of this one may have acoustic perturbations which are quantized in
terms of phonons. However, in that case, the division between the
background and linear acoustic fluctuations tends to get a bit hazy.

There is however an alternative phenomenon which does not necessitate
quantization of acoustic disturbances for it to occur, although a full
quantum description is possible in principle. This is the phenomenon
called Superresonance by us \cite{bas} which occurs in a curved
Lorentzian acoustic geometry if the acoustic spacetime is that of a
rotating black hole. Recall that such a black hole spacetime admits a
region - the ergoregion - containing the event horizon, where the time
translation Killing vector turns spacelike. The existence of this
region is crucial for the Penrose process of energy extraction from
the black hole, at the cost of its rotational energy. Superresonance
is simply an acoustic wave version of the Penrose process, wherein a
plane wave solution of a massless scalar field in the black hole
background is scattered from the ergoregion with an amplification of
its amplitude. The energy gained by the wave is at the cost of the
rotational energy of the black hole.

In the earlier paper \cite{bas}, superresonance has been shown to
occur in a certain class of analogue 2+1 dimensional rotating black
holes. Linear acoustic perturbations in such a background are shown
to scatter from the ergoregion with an enhancement in amplitude, for a
restricted  range of frequencies of the incoming wave. However, while
the plausibility of superresonance was established in that earlier
work through the analysis of  the Wronskian of the radial equation of
the acoustic perturbations, the actual frequency dependence of the
amplification factor was not worked out. In this paper this gap is
filled. We present a more detailed quantitative analysis of this
phenomenon, as also a somewhat different way of arriving at the
results of the previous paper. An explicit expression is derived for
the reflection coefficient as a function of the frequency of the
incoming wave for a certain low frequency regime. This expression is
expected to be useful for possible future experimental endeavors to
observe superresonance.

The plan of the paper is as follows: in section II, certain aspects of
the  acoustic black hole analogue we work with are reviewed. The
demonstration of superresonance is repeated with a choice of
coordinates  different from those used in the earlier work
\cite{bas}. In section III, detailed quantitative analysis of the
reflection coefficient is presented, explicitly exhibiting its
frequency dependence, albeit within a certain  restricted low
frequency regime. We conclude in section IV with a sketch of
the outlook on this class of problems.

\section{Draining Vortex Flow and superresonance}

In order to demonstrate superresonance, a certain choice is to be made
of the velocity potential of the fluid, which functions as a suitable
acoustic analogue of a Kerr black hole. In other words, the flow must
have a well-defined (ergo)region of transonic flow containing a sonic
horizon as discussed in the Introduction. In 2+1 dimensions, a
velocity profile with these properties is the `draining vortex' flow
with a sink at the origin, given by \cite{vis}
\begin{equation}
\overrightarrow{v}=-\frac{A}{r}~\hat{r}+\frac{B}{r}~\hat{\phi}~,
\label{vel}
\end{equation}
where~$ A, B $ are real and positive and $(r,\phi)$ are plane polar
coordinates. The metric corresponding to this effective geometry is ,
\begin{equation}
{ds}^2=\left(\frac{\rho_{0}}{c}\right)^{2}\left[-\left(c^2-\frac{A^2+B^2}
{r^2}\right){dt}^2+\frac{2~A}{r}~dr~dt-2B~d\phi~dt+{dr}^2+r^2~{d\phi}^2
\right]
\label{metric}
\end{equation}
A two surface, at$~r=\frac{A}{c}~$, in this flow, on which the fluid
velocity is everywhere inward pointing and the radial component of the
fluid velocity exceeds the local sound velocity everywhere, behaves as
an outer trapped surface in this acoustic geometry. This surface can
be identified with the future event horizon of the black hole. Since
the fluid velocity (\ref{vel}) is always inward pointing the
linearized fluctuations originating in the region bounded by the sonic
horizon cannot cross this boundary. As for the Kerr black hole in
general relativity, the radius of the boundary of the ergosphere of
the acoustic black hole is given by vanishing of $~ g_{00}~$,
i.e,~$r_{e}= \sqrt{A^{2}+B^{2}}/c$.  If we assume that the background
density of the fluid is constant, it automatically implies that
background pressure and the local speed of sound are also
constant. Thus we can ignore the position independent pre-factor in
the metric because it will not effect the equation of motion of
fluctuations of the velocity potential.

From the components of the draining vortex metric it is clear that the
(2+1)-dimensional curved spacetime possesses isometries that
correspond to time translations and rotations on the plane. The
solution of the massless Klein Gordon equation can therefore be
written as,
\begin{equation}
\Psi(t,r,\phi)=R(r)~e^{-i~\omega~t}~e^{i~m~\phi}
\label{psi}
\end{equation}
where $~\omega~$and$~m~$ are real and positive. In order to make
$~\Psi(t,r,\phi)~$ single valued , m should take  integer values.\\
Then the radial function $~R(r)~$ satisfies
\begin{equation}
\frac{d^{2}R(r)}{d{r}^{2}} + P_{1}(r)~\frac{dR(r)}{dr}
+Q_{1}(r)~R(r)=0~,
\label{Rr}
\end{equation}
where,
\begin{eqnarray}
P_{1}(r)=\frac{A^{2}+r^{2}~c^{2}+2~i~A~(B~m-r^{2}~\omega)}
{r~(r^{2}~c^{2}-A^{2})}\nonumber
\end{eqnarray}
and
\begin{eqnarray}
Q_{1}(r)=\frac{2~i~A~B~m-B^{2}m^{2}+c^{2}m^{2}~r^{2}
+2~B~m~\omega~r^{2}-r^{4}~\omega^{2}}{r^{2}~(r^{2}~c^{2}-A^{2})}~. \nonumber
\end{eqnarray}
Observe that the equ.(\ref{Rr}) is different from that used in \cite{bas} 
(equ. 8), since the diffeomorphisms given in equ. (5) of \cite{bas} have not 
been effected in the present case. 

Now we introduce tortoise coordinate $~r^{*}~$ through the
equation,
\begin{equation}
\frac{d}{dr^{*}}=\left(1-\frac{A^{2}}{r^{2}~c^{2}}\right)\frac{d}{dr}
\label{tor}
\end{equation}
which implies that,
\begin{equation}
r^{*}=r+\frac{A}{2~c}\log\left|\frac{r~c-A}{r~c+A}\right|
\label{tor1}
\end{equation}
This tortoise coordinate spans the entire real line as opposed to r
which spans only the half-line.The horizon at $~r=\frac{A}{c}~$ maps
to$~ r^{*}\rightarrow -\infty~$, while $~
r\rightarrow \infty~$corresponds to$~ r^{*}\rightarrow
+\infty~$ .  Let us now define a new radial function
$~G(r)$ as,
\begin{equation}
R(r)=\frac{H(r)}{\sqrt{r}}~\exp\left[\frac{i}{2}~\left\{\frac
{\omega~A}{c^{2}}~\log\left(\frac{r^2~c^2}{A^2}-1\right)-\frac{m~B}{A}~\log
\left(1-\frac{A^2}{r^2~c^2}\right)\right\}\right]
\label{Rr1}
\end{equation}
Substituting this in equation (\ref{Rr}) we observe that 
$~G(r)~$ satisfies the differential equation
\begin{equation}
\left(1-\frac{A^2}{r^2~c^2}\right)\frac{d}{dr}\left[\left(1-\frac{A^2}
{r^2~c^2}\right)\frac{d}{dr}\right]H(r)+\left[\frac{1}{c^{2}}
\left(\omega-\frac{B~m}{r^{2}}\right)^{2}-\left(1-\frac{A^2}
{r^2~c^2}\right)\left\{\frac{1}{r^{2}}\left(m^{2}-\frac{1}{4}\right)+
\frac{5~A^{2}}{r^{4}c^{2}}\right\}\right]H(r)=0
\label{Lr1}
\end{equation}
Now in terms of tortoise coordinate one obtains the modified
differential equation as,
\begin{equation}
\frac{d^{2}H(r^{*})}{d{r^{*}}^{2}}+\left[\frac{1}{c^{2}}
\left(\omega-\frac{B~m}{r^{2}}\right)^{2}-\left(1-\frac{A^2}
{r^2~c^2}\right)\left\{\frac{1}{r^{2}}\left(m^{2}-\frac{1}{4}\right)+
\frac{5~A^{2}}{r^{4}~c^{2}}\right\}\right]H(r^{*})=0
\label{lr*1}
\end{equation}
We analyze this differential equation in two distinct radial regions, viz., 
near the sonic horizon, i.e., at $r^{*}\rightarrow -\infty$ and at
asymptopia, i.e., at $~ r^{*}\rightarrow +\infty~$. In the asymptotic
region, the above differential equation can be written approximately
as,
\begin{equation}
\frac{d^{2}H(r^{*})}{d{r^{*}}^{2}}+
\frac{\omega^{2}}{c^{2}}~H(r^{*})=0
\label{Lr*2}
\end{equation}
This can be solved trivially,
\begin{equation}
H(r^{*})=R_{\omega
m}~\exp~\left(i~\frac{\omega}{c}~r^{*}\right)
+\exp~\left(-i~\frac{\omega}{c}~r^{*}\right)
\end{equation}
The first term in the above equation corresponds to reflected wave and
the second term to the incident wave, so that R is the reflection
coefficient in the sense of potential scattering. Similarly , near
the horizon the above differential equation
\label{Lr*1}can be written approximately as,
\begin{equation}
\frac{d^{2}H(r^{*})}{d{r^{*}}^{2}}+
\frac{(\omega-m~\Omega_{H})^{2}}{c^{2}}~H(r^{*})=0
\label{dLr*2}
\end{equation}
where$~\Omega_{H}~$ is the angular velocity of the sonic horizon.
We impose the physical boundary condition that of the two solutions of
this equation, only the ingoing one is physical, so that one has,
\begin{equation}
H(r^{*})=T_{\omega
m}~\exp~\left\{-i~\frac{(\omega-m~\Omega_{H})} {c}~r^{*}\right\}
\end{equation}
Now using these approximate solutions of the above differential
equation together with their complex conjugates and recalling the fact two
linearly independent solutions of this differential equation
(\ref{lr*1}) must lead to a constant Wronskian, it is easy to show
that,
\begin{equation}
1-|R_{\omega
m}|^{2}=\left(1-\frac{m~\Omega_{H}}{\omega}\right)~ |T_{\omega m}|^{2}
\label{RT}
\end{equation}
Here$~R_{\omega m}~$ and $~T_{\omega m}~$ are
amplitudes of the reflection and transmission coefficient of the
scattered wave respectively. It is obvious from the above equation
that, for frequencies in the range $~0~<~\omega~<~m~\Omega_{H}$, the
reflection coefficient has a magnitude larger than unity. This is
precisely the amplification relation that emerges in our earlier
analysis of superresonance \cite{bas}. The demonstration that the
final physical  result remains the same, despite using a different
system of coordinates in  the  acoustic spacetime, implies that the
analogue spacetime indeed exhibits the kind of general coordinate
invariance associated with standard general relativity. This relation
between reflection and transmission coefficients, however, only
establishes the {\it plausibility} of superresonance for draining vortex
flows. It is not sufficiently detailed to link up with 
possible forthcoming experimental observations. For that purpose, we
need to know the frequency dependence of both the reflection and the
transmission coefficients. This requires that we have to solve the
differential equation explicitly. In the next section we analyze
the radial differential equation in the low frequency limit
$\omega~A/c^{2}\ll 1$.

\section{The Reflection Coefficient}

We begin with defining a new function $L(r)$ through the relation
\begin{equation} 
R(r)=\frac{r~c}{A}~\exp\left[\frac{i}{2}~\left\{\frac
{\omega~A}{c^{2}}~\log\left(\frac{r^2~c^2}{A^2}-1\right)-\frac{m~B}{A}~\log
\left(1-\frac{A^2}{r^2~c^2}\right)\right\}\right]~~L(r) ~.
\end{equation}
We next introduce a new variable $~
x \equiv \frac{r^{2}~c^{2}}{A^{2}}-1~$ . Then L(x) satisfies the differential 
equation ,
\begin{equation}
\frac{d^{2}L(x)}{dx^{2}}+\left(\frac{1}{x}+\frac{1}{x+1}\right)
\frac{dL(x)}{dx}+\left[\frac{Q^{2}}{x} + \frac{1}{x+1} + (1-S^{2})+
\frac{\omega^{2}~A^{2}}{c^{4}}x\right]\frac{L(x)}{4x(x+1)}=0
\end{equation}
where,
\begin{eqnarray}
Q^{2}&=&\left(\frac{\omega~A}{c^{2}}-\frac{B~m}{A}\right)^{2}
\nonumber\\ \nonumber\\
S^{2}&=&m^{2}+\frac{2~B~m~\omega}{c^{2}}-
\frac{2~\omega^{2}~A^{2}}{c^{4}} \nonumber
\end{eqnarray}
In order to find a solution of this differential equation, we adapt a matching 
procedure employed by Starobinsky \cite{star}; for the region
\begin{eqnarray}
\left(\frac{\omega~A}{c^{2}}\right)~x ~\ll~m~~,~~ 
\frac{\omega~A}{c^{2}} ~\ll~1
\nonumber
\end{eqnarray}
the above differential equation can be approximately written as,
\begin{equation}
\frac{d^{2}L(x)}{dx^{2}}+\left(\frac{1}{x}+\frac{1}{x+1}\right)
\frac{dL(x)}{dx}+\left[\frac{Q^{2}}{x} + \frac{1}{x+1} + (1-S^{2})
\right]\frac{L(x)}{4x(x+1)}=0 ~.
\label{DLx}
\end{equation}
This is the standard Riemann-Papparitz equation with regular singular
points at $~0,-1,~$and at $\infty$. We can convert this differential
equation into the Hypergeometric form by the following substitution,
\begin{equation}
L(x)=x^{\alpha^{'}}(x+1)^{\beta^{'}} G(\overline{x})
\end{equation}
such that$~ G(\overline{x})~$ satisfies the Hypergeometric
differential equation given by
\begin{equation}
\overline{x}(1-\overline{x})\frac{d^{2}G(\overline{x})}
{d\overline{x}^{2}}+[~\gamma-(\alpha +\beta +1)\overline{x}~]
\frac{dG(\overline{x})}{d\overline{x}}-\alpha\beta G(\overline{x}) =0
\end{equation}
where
\begin{eqnarray}
\alpha^{'}&=&-~i~\frac{Q}{2}~~,~~\beta^{'}=
-\frac{1}{2}\nonumber\\ \nonumber\\
\alpha&=&-\frac{S}{2}-~i~\frac{Q}{2}~~,~~
\beta=\frac{S}{2}-~i~\frac{Q}{2}~~,~~\gamma=
1-i~Q\nonumber\\ \nonumber\\
\overline{x}&=&-x \nonumber
\end{eqnarray}
The most general solution of this differential equation is ,
\begin{equation}
G(\overline{x})=C_{1}~{}_{2}F_{1}(\alpha ,\beta ;\gamma
;\overline{x})+
C_{2}~\overline{x}^{1-\gamma}~{}_{2}F_{1}(\alpha+1-\gamma   ,\beta
+1-\gamma;2-\gamma ;\overline{x})
\end{equation}
where $~C_{1}~$and $~C_{2}~$ are arbitrary constants
to be determined.  Near the horizon, $x\sim 0$
\begin{equation}
 G(x)~\longrightarrow~C_{1}~x^{\alpha^{'}}
+C_{2}~x^{-\alpha^{'}}~.
\end{equation}
Here we have absorbed $~(-1)^{1-\gamma}~$ in $~C_{2}~$. The
first term represents ingoing wave and the second term represents
outgoing wave near the horizon. Since nothing can come out of the horizon, 
the boundary condition at the horizon implies that $~C_{2}~=0$. 
Therefore the solution of equ.(\ref{DLx}) reduces to
\begin{equation}
L(x)=C_{1}~x^{\alpha^{'}}(x+1)^{\beta^{'}}~{}_{2}F_{1}(\alpha
, \beta;\gamma ;-x)
\end{equation}

To use the matching procedure we need to know the behavior of this
solution in the asymptotic region, i.e, at$~
x~\longrightarrow~\infty~$. This can be obtained by the
transformation$~(\overline{x}\rightarrow~1/ {\overline{x}})~$
for the Hypergeometric function in above equation. Using this rule we
obtain,
\begin{eqnarray}
 L(x)&=&C_1 x^{~\alpha^{'}}~(x+1)^{\beta^{'}} \left
[ \frac{\Gamma(\gamma)~\Gamma(\beta -\alpha)}{\Gamma(\beta)~
\Gamma(\gamma-\alpha)}x^{-\alpha}~{}_{2}F_{1}\left(\alpha~ ,~\alpha +1
-\gamma ~;~\alpha +1 -\beta ~;~\frac{(-1)}{x}\right) \right . 
\nonumber\\ && \left . +x^{-~\beta}~~\frac{\Gamma(\gamma)~
\Gamma(\alpha -\beta)}
{\Gamma(\alpha)~\Gamma(\gamma-\beta)}~{}_{2}F_{1}\left (\beta~ ,~\beta
+1 -\gamma~ ;~\beta +1 -\alpha ~;~\frac{(-1)}{x}\right)
\right]\label{Lx5}
\end{eqnarray}
So in  the region $x~\gg~{Max(~m^{2},Q^{2}~)}$, we have
\begin{equation}
L(x)~\longrightarrow~C_{1}~x^{~\alpha^{'}}~(x~+~1)^{~\beta^{'}}
\left[\frac{\Gamma(\gamma)~\Gamma(\beta -\alpha)}{\Gamma(\beta)~
\Gamma(\gamma-\alpha)}~x^{-\alpha} +\frac{\Gamma(\gamma)~
\Gamma(\alpha -\beta)}{\Gamma(\alpha)~\Gamma(\gamma-\beta)}~
x^{-\beta}\right]
\label{Lx4}
\end{equation}
Hence the approximate form of $~L(x)~$ in the region,
\begin{eqnarray}
{Max(~m^{2},Q^{2}~)}~\ll~x~\ll~m\Big{/}\left(\frac{\omega~A}{c^{2}}
\right)\nonumber
\end{eqnarray}
is given by
\begin{eqnarray}
L(x)&=&\frac{C_{1}}{\sqrt{~x~}}~\left[\frac{\Gamma(\gamma)
~\Gamma(\beta -\alpha)}{\Gamma(\beta)~
\Gamma(\gamma-\alpha)}~x^{\frac{S}{2}} +\frac{\Gamma(\gamma)~
\Gamma(\alpha -\beta)}{\Gamma(\alpha)~\Gamma(\gamma-\beta)}~
x^{-\frac{S}{2}}\right]\nonumber\\ \nonumber\\ \nonumber\\
&=&\frac{1}{\sqrt{~x~}}\left[~D_{1}~x^{\frac{S}{2}}~+~D_{2}~
x^{-\frac{S}{2}}\right] \label{Lx3}
\end{eqnarray}
where,
\begin{eqnarray}
D_{1}&=&\frac{\Gamma(\gamma) ~\Gamma(\beta
-\alpha)}{\Gamma(\beta)~ \Gamma(\gamma-\alpha)}~C_{1}~\nonumber\\
D_{2}&=&\frac{\Gamma(\gamma)~\Gamma(\alpha -\beta)}
{\Gamma(\alpha)~\Gamma(\gamma-\beta)}~C_{1}~\nonumber
\end{eqnarray}

The solution (\ref{Lx3}) is in a form where it is not obvious which
combinations of the complex coefficients $D_1,D_2$ correspond to the
ingoing and outgoing modes, This information can be retrieved from the
energy flux per unit time radially across an arbitrarily chosen
surface at a constant radial coordinate $r$ in the region $~
Max(~m^{2},Q^{2}~)~\ll~x~\ll~m\Big{/}\left( \frac{\omega~A}{c^{2}}
\right)$. This is obtained by the surface integral
\begin{eqnarray}
FL&=&-\int_{0}^{~2~\pi} T_{\mu\nu}~l^{\mu}~k^{\nu} \sqrt{-g}~d\phi\nonumber\\
&=&~i~\pi~S~\omega~\Big(D_{1}^{*}~D_{2}-D_{1}~D_{2}^{*}\Big)\nonumber\\
&=&\frac{\pi~\omega~S~}{2}
~\Big(|D_{1}+i~D_{2}|^{2}-|D_{1}-i~D_{2}|^{2}\Big)
\end{eqnarray}
where $ T_{\mu\nu}$ is the energy momentum tensor for $ \Psi $, $~l^{\mu}~$ is the normal to a constant r surface ,  $~k^{\mu}~$ is the time translational killing vector and g is the determinant of the acoustic metric.\\
Given that the outgoing mode of the wave corresponds to a positive value of 
radial component of the flux, and the ingoing mode just the opposite, one 
concludes that the amplitudes of the incident and reflected wave are 
proportional to $~(D_{1}~-~i~D_{2})~$ and 
$~ (D_{1}~+~i~D_{2})~$ respectively. 

One can now write down an explicit expression for the amplification
factor(AF); this is given by,
\begin{eqnarray}
AF&~\equiv~&1-|R_{\omega
m}|^{2}=1-\left|\frac{D_{1}+iD_{2}}
{D_{1}-iD_{2}}\right|^{2}\nonumber\\ \nonumber\\
&~=~&\frac{2~i~(D_{1}~D_{2}^{*}~-~D_{1}^{*}~D{_2})}{\left|D_{1}
-~i~D_{2}\right|^{2}}\nonumber\\ \nonumber\\
&~=~&\frac{2~Q~|C_{1}|^{2}}{S~\left|D_{1}-i~D_{2}\right|^{2}}=
\frac{2~\omega~A}{S~c^{2}}\frac{|C_{1}|^{2}}{~\left|D_{1}-i~D_{2}
\right|^{2}}~\left(1-\frac{m~\Omega_{H}}{\omega}\right) \label{Rf}
\end{eqnarray}
It follows from this that the reflection coefficient exceeds unity in the 
frequency range $ ~0~<~\omega~<m~\Omega_{H}~$. Comparing this  with 
the equation (\ref{RT}), we can identify$~ T_{\omega m}~$ in this
frequency range as,
\begin{equation}
|T_{\omega m}|^{2}=\frac{2~\omega~A}{S~c^{2}}\frac{
|C_{1}|^{2}}{~\left|D_{1}-i~D_{2}\right|^{~2}}
\label{Trm}
\end{equation}

Equations (\ref{Rf}-\ref{Trm}) exhibit the frequency dependence of
both the reflection and transmission coefficients and hence of the
amplification factor in the low frequency regime. If vortex motion
with a sink is realized in the laboratory, our result can be used to
provide an order of magnitude estimate for the amplification
factor. According to our assumption these relations are valid in the
frequency range where$~\frac{\omega~A}{c^{2}} ~\ll~1 ~$ , i.e, when wave
length of the sound wave is much larger than the radius of the
horizon. Clearly, this frequency range does not cover the entire
superresonant range $~0~$ to$~~m~\Omega_{H}~~$. It is interesting and
of importance to find the behavior of the reflection coefficient near
the critical point around $\omega=~m~\Omega_{H}~~$ where the
reflection coefficient is expected to cross unity. We hope to report
on this in the near future.

\section{Outlook}

The link with possible experimental observation of superresonance is
yet incomplete, despite the explicit expressions we have derived for
the reflection and transmission coefficients. One needs to input into
these results typical characteristics of draining vortex flows in
fluids like superfluid helium which obey our assumptions of
barotropicity, irrotationality and zero viscosity. These
characteristics are to be derived from the macroscopic quantum
mechanics of superfluid helium, and novel features like flux
quantization are expected to play a non-trivial role. In particular,
flux quantization may well lead to the phenomenon of {\it discrete}
amplification discussed in \cite{bas}, which would perhaps facilitate
observation. We should mention that ergoregions in superfluid helium
have been considered in ref. \cite{vol}, \cite{fis}, although from a
somewhat different standpoint. These authors do not consider vortex
flows with a drain, and as such, their acoustic spacetime does not
share all characteristics of black holes, like an outer trapped
surface. 

Another feature which we have ignored in the foregoing analysis is the
existence of shock fronts at the interface of normal and transonic
flows.  Indeed, in astrophysical black holes, analysis of these shock
fronts constitutes an important component of models used in
observational work on these objects. The existence of these shock
fronts are in part responsible for the non-observation of superradiant
scattering in such black holes. The issue of shocks in acoustic black
hole analogues thus assumes special significance for future work on
this topic.

One of us (PM) thanks J. Bhattacharjee, G. Bhattacharya and
A. Chatterjee for helpful discussions.

\end{document}